\documentclass[review,a4paper,3p,12pt]{elsarticle}

\usepackage{svg}
\usepackage[T1]{fontenc}
\usepackage[utf8]{inputenc}
\usepackage{amsmath,amssymb}
\usepackage[colorlinks=true,citecolor=ForestGreen,linkcolor=Red,urlcolor=Blue,hypertexnames=false]{hyperref}
\usepackage{graphicx}
\usepackage{graphics}
\usepackage{natbib}
\usepackage{multirow}%
\usepackage{booktabs}
\usepackage{threeparttable}

\begin{document}

\begin{frontmatter}
\title{A simple model of decision-making in the application process}
\author[inst2,inst7]{Fanyuan Meng}
\author[inst1,inst3,inst7]{Hui Xiao}

\author[inst1,inst3]{Xinlin Wu}
\author[inst1,inst3,inst4]{Xiaojun Hu}
\author[inst5]{Xiaojie Niu}
\author[inst6,inst8]{Sheng Chen}
\author[inst1,inst3,inst8]{Yu Liu}

\affiliation[inst1]{organization={Department of Systems Science, Faculty of Arts and Sciences},
    addressline={Beijing Normal University}, 
    city={Zhuhai},
    postcode={519087},
    country={China}}

\affiliation[inst2]{organization={Research Center for Complexity Sciences},
    addressline={Hangzhou Normal University}, 
    city={Hangzhou},
    postcode={311121}, 
    opinion={Zhejiang},
    country={China}}

\affiliation[inst3]{organization={International Academic Center of Complex Systems},
    addressline={Beijing Normal University}, 
    city={Zhuhai},
    postcode={519087},
    country={China}}

\affiliation[inst4]{organization={School of Systems Science},
    addressline={Beijing Normal University}, 
    city={Beijing},
    postcode={100875},
    country={China}}

\affiliation[inst5]{organization={Faculty of Education},
    addressline={Beijing Normal University}, 
    city={Beijing},
    postcode={100875},
    country={China}}

\affiliation[inst6]{organization={Research Center for Mathematics},
    addressline={Beijing Normal University}, 
    city={Zhuhai},
    postcode={519087},
    country={China}}

\affiliation[inst7]{These authors contribute equally.}
\affiliation[inst8]{Corresponding authors: shengchen@bnu.edu.cn;  yu.ernest.liu@bnu.edu.cn.}


\begin{abstract}
In decision-making, individuals often rely on intuition, which can occasionally yield suboptimal outcomes. This study examines the impact of intuitive decision-making on individuals who are confronted with limited position information in the job application process. We propose a measure, the mismatch index, that gauges allocation efficiency by comparing the final application rate to the preset admission rate. By simulation and analytical results, we counter-intuitively find that under the intuitive strategy, acquiring more information does not always lead to more efficient allocation. Additionally, a shift from despondency to a bandwagon effect occurs when the initial application rate surpasses the admission rate, which can be observed in our field experiments. Meanwhile, experimental data also unveil variations in individuals' reliance on intuition, indicating the presence of inherent adventurous and conservative inclinations. To account for these effects, we introduce an enhancement factor into our model. The improved results align well with these real data, showing that compared to mediate competitive scenarios, individuals exhibit a stronger conservative tendency in fierce or less competitive scenarios. These findings offer significant insights into resource allocation, especially in the competitive job market context.
\end{abstract}


\end{frontmatter}

\section{Introduction}
\label{sec:introduction}

The exploration of decision-making processes has witnessed an unprecedented surge, particularly within the dynamic and intensely competitive landscapes that define various facets of our society. This heightened interest emphasizes the growing recognition of the pivotal role that decision-making plays, not only in shaping individual trajectories but also in influencing broader societal outcomes. From the cutthroat world of college admissions, where aspiring students vie for coveted spots in prestigious institutions \cite{rigol2003admissions, mengash2020using, humlum2017college}, to the fiercely competitive landscape of job applications \cite{jones2006recruiting, deming2021growing}, where candidates strive to distinguish themselves in crowded markets, the dynamics of decision-making have become a central focus in disciplines including economics, psychology, and sociology. 

The study of decision-making in competitive contexts has transcended mere examinations of strategic considerations and aims to reveal how individuals navigate intricate scenarios through a nuanced blend of psychological processes. These mechanisms, deeply entrenched in cognitive and non-cognitive processes, exert a profound influence on individuals as they traverse the complex terrain of competitive decision-making \cite{lant2002information, van2021modeling}.

Cognitive biases, epitomizing the nuanced complexities of human psychology, shed light on the intricate challenges individuals encounter when simplifying decision-making through mental shortcuts or heuristics \cite{berthet2021measurement, acciarini2021cognitive, haselton2015evolution, meng2022whom}. One prominent cognitive bias, confirmation bias, stands out as a pervasive and impactful force in the decision-making landscape. This bias manifests when individuals selectively seek, interpret, and remember information that aligns with their pre-existing beliefs or convictions, potentially leading to the formation of biased strategies in the context of competitive decision-making \cite{klayman1995varieties, glick2017believing, nickerson1998confirmation}. Additionally, the influence of confirmation bias extends beyond individual decision-makers, impacting the strategies of entire organizations and contributing to the perpetuation of suboptimal or outdated approaches \cite{ofem2023decision}.

Another type of cognitive bias, the availability heuristic, adds another layer of complexity in competitive decision-making. This cognitive shortcut involves making decisions based on readily available information or examples that come to mind easily, often leading individuals to overestimate the likelihood of events that are easily recalled \cite{folkes1988availability, macleod1992memory}. In the competitive landscape, individuals may rely on recent events or information derived from a limited sample size that inadequately represents the broader population. This tendency can lead to the fallibility of decision-makers, as they draw conclusions from insufficient information, potentially overlooking less prominent but equally important factors \cite{dale2015heuristics, tversky1973availability}. Consider, for instance, a manager making decisions about resource allocation based on recent success stories within the industry. Relying solely on the availability heuristic, they might overlook the diverse challenges faced by different companies, potentially leading to suboptimal resource distribution \cite{dube1988availability}. In the context of college admissions, admissions officers might be influenced by recent trends in student applications, possibly giving undue weight to these trends while inadvertently overlooking the unique qualities of individual candidates and their potential contributions to the academic community \cite{smith1994cognitive}. Students face similar issues, often referred to in Chinese social media or news as ``large-small-year'' (literally translated). This term means that the application-to-admission ratios of some universities display a wave-like pattern over the years, with one year experiencing a low ratio, followed by a year with a much higher ratio that significantly exceeds expectations. Thus recognizing and mitigating these biases become imperative for individuals and organizations striving to make sound and objective decisions \cite{bieske2023trends, featherston2019interventions}. Strategies that promote awareness, diverse information sources, and critical reflection can serve as effective countermeasures to these cognitive pitfalls. By delving into the nuanced interplay of cognitive biases in competitive decision-making, our research aims to contribute not only to the theoretical understanding of these biases but also to offer practical insights that can enhance decision-making processes across various domains \cite{baddeley2004introduction, saposnik2016cognitive}. 

Furthermore, the non-cognitive process emerges as another fundamental and intricate component, exerting a pronounced influence, especially in competitive scenarios \cite{lerner2015emotion, schwarz2000emotion, lerner2000beyond}, e.g., the emotional state of an individual significantly shapes how they perceive and assess risks, thereby intricately interweaving the fabric of decision-making dynamics \cite{loewenstein2001risk, slovic2006risk}. Understanding these affective nuances becomes crucial, as they play a pivotal role in shaping the strategies individuals adopt within competitive environments \cite{simon1987making}. 

One aspect is that, elevated levels of fear or anxiety can significantly impact decision-making \cite{da2023cognitive}, often prompting a more conservative approach \cite{kahneman2013prospect, hartley2012anxiety, wagner2019anxiety}. In the face of uncertainty or perceived threats, individuals tend to prioritize risk aversion, seeking to minimize potential negative outcomes \cite{maner2007dispositional}. This heightened vigilance, while potentially preventing undue risks, may also hinder innovation or bold decision-making that could lead to significant gains \cite{kimball1993standard, brandstatter2006priority}. Therefore, the delicate balance between risk aversion and the pursuit of opportunities becomes a critical consideration for individuals navigating competitive landscapes. Conversely, a state of heightened excitement or overconfidence introduces a different dynamic into the decision-making equation \cite{loewenstein2001risk, sharot2016forming}. In such scenarios, individuals may display an increased tolerance for risk, potentially leading to more audacious decisions \cite{sharot2023and, jordan2011something, march1987managerial}. The allure of potential rewards and a sense of invincibility can drive individuals to venture into uncharted territories. While this boldness may result in groundbreaking innovations or strategic moves, it also carries the inherent risk of overlooking potential pitfalls and underestimating challenges \cite{christensen2013disruptive}.

However, prior endeavors have often overlooked the intricate interplay of psychological mechanisms, specifically the integration of cognitive processes in the context of decision-making within competitive environments. Addressing this gap forms the foundation of our work, where we develop a new model designed to capture the subtleties inherent in decision-making within the application processes. Through analytical, simulation, and experimental analyses, we offer a thorough examination of the effects of insufficient sample size (cognitive bias) on human resource allocation. In addition, we reveal that in most cases, a naive strategy of decision-making is adopted by the majority of the population, but risk appetite and aversion both exist in the real world. To improve the predictive accuracy, we introduce an enhancement factor by integrating two parameters characterizing adventurous and conservative inclination (non-cognitive bias) into our model.

\section{Model}

In our model, we imagine an organization aims at hiring a specific number $n_a$ of individuals from a targeted population with a total size of $N$. Therefore, the \emph{admission rate} can be denoted by $p_a=n_a/N$. Initially, we assume that a certain number of $n_0$ of individuals have the inclination to apply for the position, resulting in an \emph{initial application rate} of $p_0=n_0/N$. All individuals will then attempt to gather information to aid in their decision-making. Specifically, each individual evaluates the degree of competition by the \emph{perceived application rate} $p_s=r/(R+1)$; that is, they consult random $R$ individuals (excluding themselves) to find that the number of applicants is $r$ (including themselves). Note that consulting an insufficient subset of the total population, i.e. $R < N-1$, can significantly skew evaluations of the true value of $p_0$. This can lead to phenomena like the majority illusion \cite{lerman2016majority} or perception bias \cite{lee2019homophily, yu2023opinion}. Specifically, when $p_s > p_0$, individuals tend to overestimate the fraction of applicants. On the other hand, when $p_s < p_0$, they tend to underestimate this fraction. It is further assumed that individuals would employ an intuitive strategy when making decisions: If the perceived application rate $p_s$ exceeds the admission rate $p_a$, i.e., $p_s \geq p_a$, the individual would feel that the competition is too intense and finally decide to give up the application; otherwise, the applicant proceeds with the application. 

The number $R$ of random consultations and intuitive decision-making strategy significantly influence the final application rate, represented as $p = n/N$, where $n$ stands for the final number of applicants. Furthermore, it impacts the efficiency of resource allocation, which can be quantified by the difference between the final application rate $p$ and the admission rate $p_a$, referred to as the \emph{mismatch index} $m$:
\begin{equation}
m = p - p_a,
\end{equation}
where $m$ falls within the range $[-1, 1]$. A mismatch index value of 0 indicates the most efficient resource allocation, meaning that the admission and application numbers of individuals are perfectly balanced. If $m \neq 0$, it suggests a certain degree of resource inefficiency, potentially leading to resource waste.

\section{Results}
To facilitate understanding, we have collected the parameters and variables involved in the aforementioned model as follows:
\begin{itemize}
    \item $N$: the total population size.
    \item $n_a$: the number of available positions. 
    \item $p_a$: the admission rate, i.e., $p_a=n_a/N$.
    \item $n_0$: the number of initial applicants.
    \item $p_0$: the initial application rate, i.e., $p_0=n_0/N$.
    \item $R$: the number of random consultations.
    \item $r$: the number of applicants encountered during $R$ consultations.
    \item $p_s$: the perceived application rate, i.e., $p_s = r/(R+1)$.
    \item $n$: the final number of applicants. 
    \item $p$: the final application rate, i.e., $p=n/N$.
    \item $m$: the mismatch index, i.e., $m=p-p_a$.
\end{itemize}

\subsection{Analytical Results}

To assess the final application rate, we consider two scenarios:

(1) For the scenario where the individual has the initial inclination to apply, we calculate the probability $p^+(r, R)$ that the individual finds $r$ individuals who have the initial inclination to apply after conducting $R$ consultations. This probability is given by
\begin{equation}
    p^+(r, R) = p_0 \frac{C_{n_0-1}^{r-1} C_{N-n_0}^{R-r+1}}{C^{R}_{N-1}}, \quad r = 1,2,\ldots,R.
    \label{eq:p_plus}
\end{equation}
The term $C_{n_0-1}^{r-1}$ represents the number of ways to choose $r-1$ individuals from the $n_0-1$ who have the initial inclination, and $C_{N-n_0}^{R-r+1}$ represents the number of ways to choose $R-r+1$ individuals from the remaining $N-n_0$ individuals who do not have the initial inclination. The denominator $C^{R}_{N-1}$ denotes the total number of ways to choose $R$ individuals from the entire population excluding the individual.

(2) For the scenario where the individual does not have the initial inclination to apply, we calculate the probability $p^-(r, R)$ that the individual finds $r$ individuals who have the initial inclination to apply after conducting $R$ consultations. This probability is given by
\begin{equation}
    p^-(r, R) = (1-p_0) \frac{C_{n_0}^{r} C_{N-1-n_0}^{R-r}}{C^{R}_{N-1}}, \quad r = 0,1,\ldots,R.
    \label{eq:p_minus}
\end{equation}
The term $C_{n_0}^{r}$ represents the number of ways to choose $r$ individuals from the $n_0$ individuals who have the initial inclination, and $C_{N-1-n_0}^{R-r}$ represents the number of ways to choose $R-r$ individuals from the remaining $N-1-n_0$ individuals who do not have the initial inclination. Again, the denominator $C^{R}_{N-1}$ denotes the total number of ways to choose $R$ individuals from the entire population excluding the individual.

Combined with the above two scenarios, the probability that the individual finally decides to apply for the position can be obtained by
\begin{equation}
p(r,R) = 
    \begin{cases}
        p^+(r,R) + p^-(r,R), \quad r \geq 1, \\
        p^-(r,R), \quad r = 0.
    \end{cases}
\end{equation}
Finally, we calculate the expected final application rate $\overline{p}$ by summing the probabilities for all $r$ as follows
\begin{equation}
\overline{p} = \sum_{r=1}^{\lfloor p_a (R+1) \rfloor} p^+(r,R) + \sum_{r=0}^{\lfloor p_a (R+1) \rfloor} p^-(r,R).
\label{eq:final_P}
\end{equation}

Here we discuss specific scenarios that arise under particular conditions as follows: When $p_a=0$ or $p_a=1$, the outcomes are well-defined. On the one hand, if $p_a=0$, it indicates that no individuals will be hired, and then all individuals in the population will certainly forego the application, i.e., $p=0$. On the other hand, if $p_a=1$, it implies that all individuals will be hired, prompting every individual to apply, i.e., $p=1$. Thus under these two cases, the resource allocation is the most efficient, i.e., $m=0$. 

When $R=N-1$, the final application rate $p$ is solely determined by the difference between $p_0$ and $p_a$. If $p_0 \geq p_a$, suggesting that the initial inclination application rate is equal to or greater than the admission rate, the entire population will abstain from applying for the position, resulting in $p=0$. In this case, the mismatch index is $m=-p_a$ (lower bound), reflecting a pessimistic outcome: a higher admission rate $p_a$ implies a poor resource allocation. Conversely, if $p_0 < p_a$, indicating that the initial inclination application rate is lower than the admission rate, the entire population will opt to proceed with the application, leading to $p=1$. Consequently, the mismatch index becomes $m=1-p_a$ (upper bound), implying fiercer competition. This indicates that only when there is a lower admission rate $p_a$, will the allocation of resources be efficient.

\subsection{Simulation Results}
\begin{figure}[!ht]
    \centering
    \includegraphics[width=\linewidth]{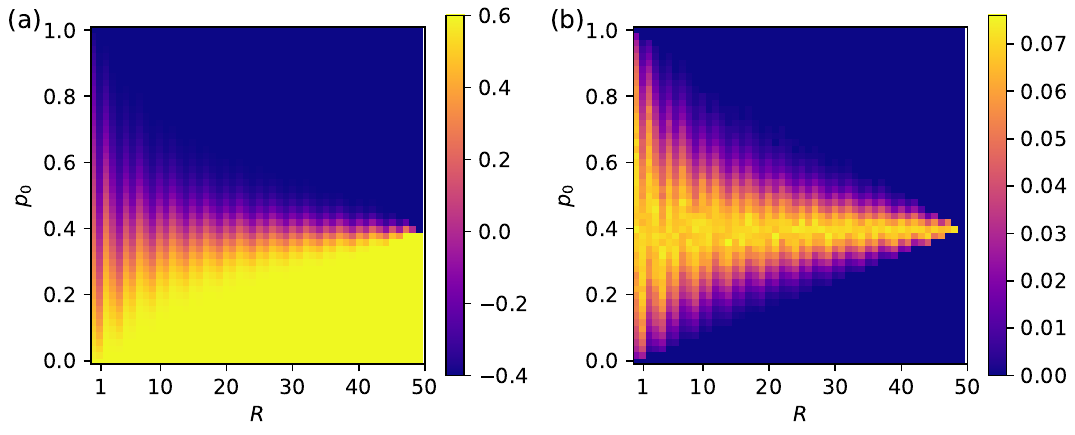}
    \caption{The heatmaps depict the mismatch index $m$ in relation to the number $R$ of consultations and the initial application rate $p_0$. Panel (a) presents the average mismatch index $m$, while panel (b) displays the standard deviation of $m$. The simulation was conducted with $N=50$, $p_a=0.4$, and a total of $500$ independent implementations.}
    \label{fig:R_p0}
\end{figure}

We demonstrate the impact of the admission rate $p_a$, the initial application rate $p_0$, and the number $R$ of random consultations on the mismatch index $m$ by the heatmaps. From Fig. \ref{fig:R_p0}(a), it is evident that for fixed admission rate (e.g., $p_a=0.4$), an extremely higher or lower initial application rate $p_0$ leads to either extreme pessimism or fierce competition, both of which results in inefficient resource allocation regardless of the value of $R$. Furthermore, a larger value of $R$ implies that the higher efficiency area becomes smaller, which suggests that the individuals' decision-making becomes increasingly polarized. This trend is confirmable by the standard deviations in Fig.~\ref{fig:R_p0}(b). Such a trend can be detrimental to resource allocation.

\begin{figure}[!ht]
    \centering
    \includegraphics[width=\linewidth]{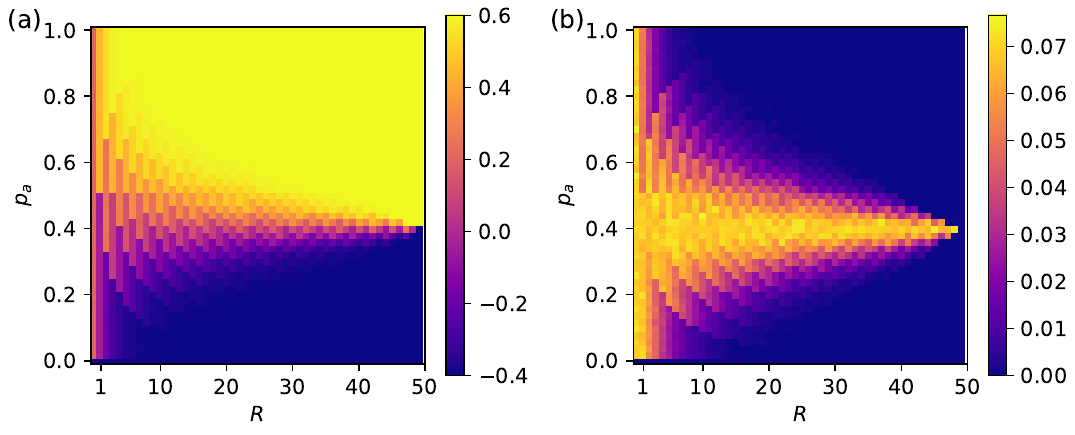}
    \caption{The heatmaps depict the mismatch index $m$ in relation to the number $R$ of consultations and the admission rate $p_a$. Panel (a) presents the mismatch index $m$, while panel (b) displays the standard deviation of $m$. The simulation was conducted with $N=50$, $p_0=0.4$, and a total of $500$ independent implementations.}
    \label{fig:R_pa}
\end{figure}

In addition, in Fig. \ref{fig:R_pa}, it is clearly observed that the effect of admission rate $p_a$ and initial application rate $p_0$ on the mismatch index $m$ are reversed. Specifically, for fixed initial application rate (e.g., $p_0=0.4$), an extremely higher or lower admission rate $p_a$ leads to either fierce competition or extreme pessimism, both of which results in inefficient resource allocation regardless of the value of $R$. Additionally, a larger number $R$ of consultations also causes the higher efficiency area to shrink as well. This trend is confirmable by the standard deviations in Fig. \ref{fig:R_pa}(b). This trend is also detrimental to resource allocation.

\newpage
To summarize, first of all, the intuitive strategy mentioned above is very easy to think of and, at first glance, appears to be a good choice. It might also be the strategy that many people tend to adopt. However, if everyone in society adopts this naive strategy, then the more information gathered (namely with larger $R$), the more likely it is to lead to poor decisions. This can result in unnecessary intense competition where it should not exist, or in a collective negative mindset that results in positions that should attract applicants being left unattended. Both of these effects are detrimental to resource allocation. The effort to gather more information is only useful if the strategy is good; otherwise, acquiring more information can lead us further in the wrong direction.

\subsection{Experimental Results and Comparisons}
To verify our model, we conducted field experiments. The experiments were conducted during the graduation season, when students had to decide whether to compete for a specific position based on the information they gathered. Participants were assigned points based on their choices: 0 point for not participating, $+1$ point for successfully getting the position, and $-1$ point for participating but failing to get the position. The objective of the students in the experiment was to maximize their scores. 

\begin{figure}[!ht]
    \centering
    \includegraphics[width=0.8\linewidth]{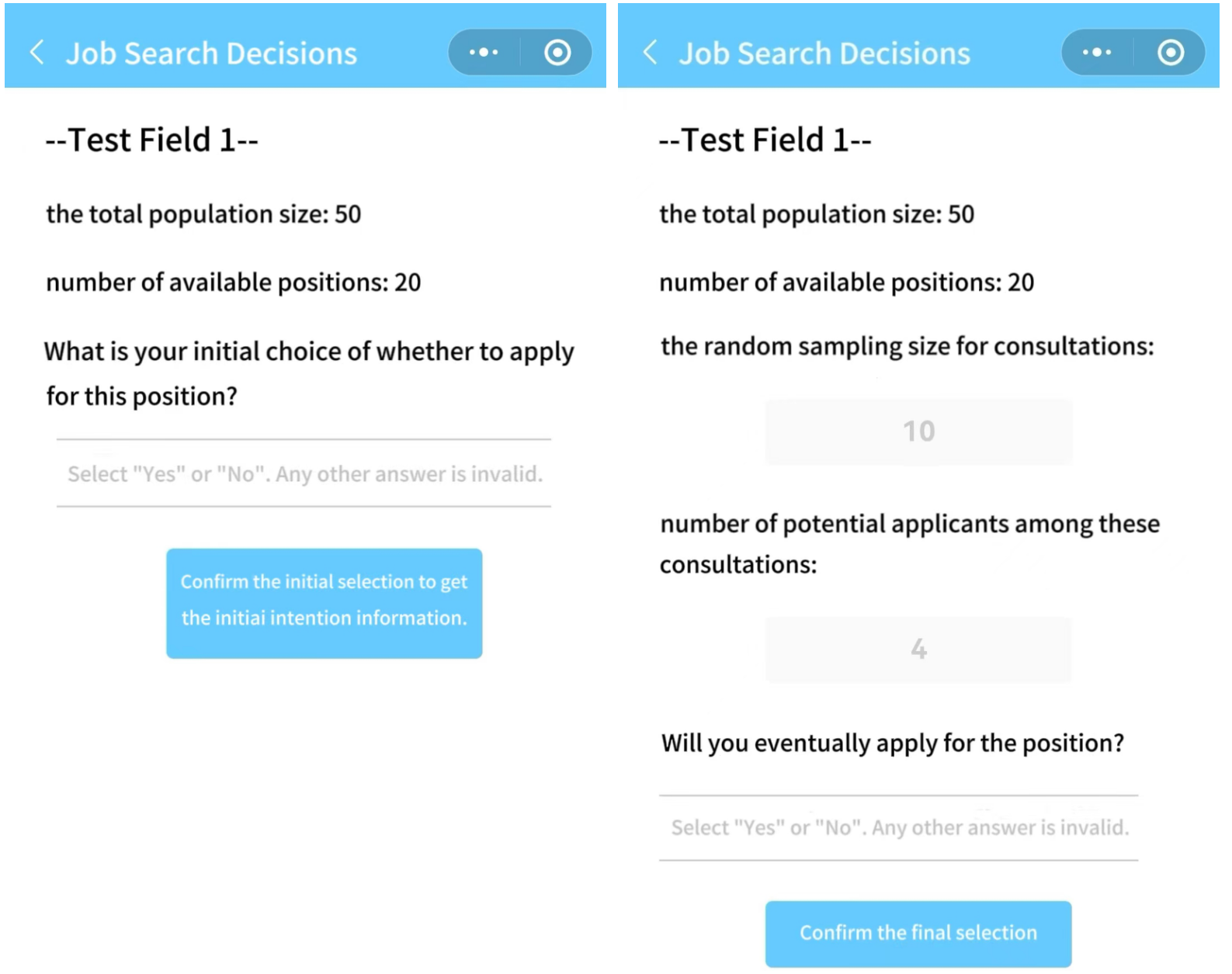}
    \caption{The illustration of the WeChat mini-program used in the experiment.}
    \label{fig:wechat}
\end{figure}

To ensure that the collected data reflected the genuine psychological thoughts of the participants when making decisions, appropriate training was provided prior to the formal experiment. The experiment was conducted using a WeChat mini-program specifically developed for this purpose. The formal experiment proceeded in two steps: (1) Participants were first presented with a page with information on the total population size ($N$) and the number of available positions ($n_a$), prompting them to make a preliminary decision on whether to apply for the position (see the left panel of Fig. \ref{fig:wechat}); (2) Once finished, participants were immediately directed to the next page which displayed additional information: first, the number $R$ of random consultations the participants conducted, and second, the number $r$ of people who had the initial inclination to apply among these consultations (assuming that this data was obtained by the participants themselves through inquiries with others). Based on this information, participants could evaluate the degree of competition to make the final decision. 

The experiment comprised eight rounds, with each round involving the participation of 50 students from Beijing Normal University. The complete data of the results of our experiments can be found in the Supplementary Table, and here we summarize the results shown in Table \ref{tab:exp}. The number of available positions ($n_a$) varied for different cases, presenting the \emph{Normal} (Cases 1-5, $n_a=20$), \emph{Hard} (Cases 6-12, $n_a=5$ and $n_a=10$), and \emph{Easy} (Cases 13-15, $n_a=30$) levels of competition.
\begin{table}[!ht]
\centering
\footnotesize
\begin{threeparttable}
\caption{Summary of field experimental, analytical, and improved results.}
\label{tab:exp}
\begin{tabular}{c | rrr | rrrr  r l r}
\toprule
Case & $n_a$ & $n_0$ & $R$ & $+1$ & $0$ & $-1$ & $m$ & $m^*$  & &  $m^{**}$\\
\midrule
1 & 20 & 10 & 15   & 20.0 & 6.5 & 23.5  & 0.47     & 0.59 &  & 0.46 \\
2 & 20 & 10 & 25   & 20.0 & 6.8 & 23.3  & 0.47     & 0.60 &  & 0.46\\
3 & 20 & 20 & 25   & 20.0 & 15.9 & 14.1 & 0.28     & 0.12 &  & 0.31\\
4  & 20 & 40 & 15   & 18.3 & 30.0 & 1.8  & 0        & $-0.40$ & $\dag$ & $-0.03$\\
5  & 20 & 40 & 25   & 17.6 & 31.6 & 0.8  & $-0.03$  & $-0.40$ & $\dag$ & 0.01 \\
\midrule
6 & 5 & 3 & 15     & 5.0 & 17.6 & 27.4 & 0.55      & 0.66 & & 0.56\\
7 & 5 & 5 & 15     & 5.0 & 19.0 & 26.0 & 0.52      & 0.38 & & 0.51\\
8  & 5 & 10 & 15     & 5.0 & 25.5 & 19.5 & 0.39      & $-0.01$ & $\dag$ &0.39 \\
9  & 5 & 15 & 15     & 5.0 & 32.0 & 13.0 & 0.26      & $-0.09$ & $\dag$ &0.26 \\
10 & 10 & 5 & 15    & 10.0 & 11.5 & 28.5 & 0.57     & 0.77 & &0.53\\
11 & 10 & 10 & 15   & 10.0 & 18.3 & 21.8 & 0.44     & 0.40 & & 0.40\\
12  & 10 & 20 & 15     & 10.0 & 27.1 & 12.9 & 0.26      & $-0.16$ & $\dag$ & 0.16\\
\midrule
13 & 30 & 20 & 25     & 30.0 & 6.6 & 13.4 & 0.27      & 0.40 & & 0.28\\
14  & 30 & 30 & 25     & 30.0 & 13.0 & 7.0 & 0.14    & $-0.12$ & $\dag$ & 0.10\\
15  & 30 & 40 & 25     & 25.8 & 24.1 & 0.1 & $-0.08$ & $-0.60$ & $\dag$ & $-0.07$\\
\bottomrule
\end{tabular}
\begin{tablenotes}[flushleft]
\footnotesize
    \item[~]The column ``$+1$'' indicates the number of people with a score of $+1$. The same applies to the columns ``0'' and ``$-1$''. The column ``$m$'' is the mismatch index calculated from the experiments, while the column ``$m^*$'' is the mismatch index analytically calculated. In addition, ``$m^{**}$'' represents the value of the mismatch index calculated by Eq.~\eqref{eq:improve_m} in the improved model with adventurous and conservative behaviors.
\end{tablenotes}
\end{threeparttable}
\end{table}

We can observe that: (1) In the \emph{Normal} case, that is, when there is a middle level of available positions ($n_a = 20$), the experimental and the analytical results match very well (as indicated by the small difference in the values of $m$ and $m^*$), except for the 4th and 5th cases (marked with $\dag$), in which cases, $n_0 = 40$, indicating an anticipated fierce competition. Under fierce competition, intuitively, most people are reluctant to apply. As a result, the number of individuals scoring 0 would be significantly higher than those scoring either $+1$ or $-1$. This is reflected in the analytical result $m^* = -0.4 \ll 0$, that is, the most extreme situation where all individuals ultimately chose not to apply. However, in the field experiments, the results were not as extreme; nevertheless, it was indeed observed from the data that the number of individuals scoring 0 (which is $30.0$ and $31.6$) was significantly higher than the other three cases.

(2) In the \emph{Hard} case, the actual results and the simulated data are generally consistent in all cases except for the 8th, 9th, and 12th cases. Likewise, in these inconsistent cases, $n_0$ is exceptionally large. The reason is similar to what was mentioned above: people anticipate intense competition. Nevertheless, in real situations, there are still some who are willing to take the risk, resulting in outcomes that are not as extreme as predicted.

(3) In the \emph{Easy} case, a similar trend follows. In the 14th and 15th cases, where, again, the competition is extremely intense, experimental and analytical results did not match as well.

To summarize, the reason for the discrepancies in some specific cases is likely due to the fact that when $p_0 > p_a$ or $p_0 \gg p_a$, some individuals in the group still retain the courage to face the competition or have a strong desire for the position. These emotions or feelings are also influencing factors that affect the decision-making behavior of individuals. In the following subsection, we analyze and explore these types of behaviors (non-cognitive factors).

\subsection{Adventurous and Conservative Behaviors}
\begin{figure}[!ht]
\centering
    \includegraphics[width=\linewidth]{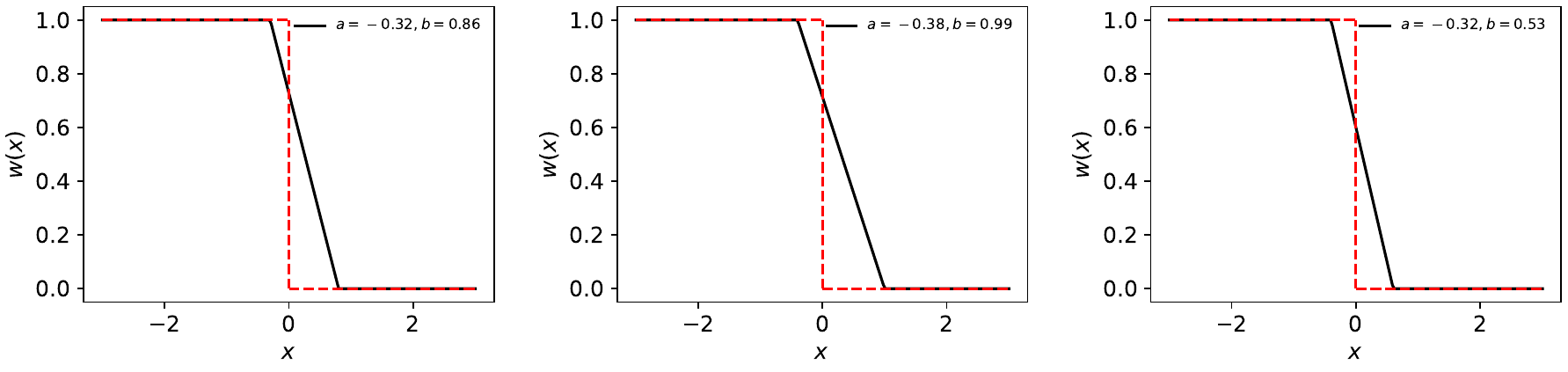}
      \caption{The figures depicted below show the optimal values of parameters $a$ and $b$ for varying levels of competition based on data obtained from field experiments. The panels, from left to right, represent the \emph{Easy}, \emph{Normal} and \emph{Hard} cases, respectively. In our methodology, the determination of the optimal values for $a$ and $b$ involves a numerical minimization process of the associated error function. We systematically substitute a wide range of parameter values for $a$ and $b$ into the error functions and identify the point where the error function reaches its minimum, thereby leading to the selection of the optimal parameter values.}
      \label{fig:opitimal}
\end{figure}

Building upon the preceding discourse, it becomes evident that relying solely on intuitive decision-making, characterized by criteria such as ``if $p_s< p_a$, then apply'' or ``if $p_s \geq p_a$, then drop the application'', falls short in capturing the nuances of decision-making tendencies in the real world. Hence, we advocate for the incorporation of another factor to modulate the terms $p^+(r,R)$ and $p^-(r,R)$ in Eqs. \eqref{eq:p_plus} and \eqref{eq:p_minus}. This factor, denoted as $w(x)$, can be constructed as follows:
\begin{equation}
w(x)= 
\begin{cases}
1 & \quad x<a, \\
1+k(x-a) & \quad a \leq x \leq b, \\ 
0 & \quad x>b.
\end{cases}
\end{equation}
In this context, $k \equiv 1/(a-b) < 0$, and $x$ is represented as $x = p_s-p_a = (r/R) - (n/N)$. Additionally, the parameters $a$ and $b$ collaboratively govern the spectrum of conservative and adventurous inclinations, as illustrated in Fig. \ref{fig:opitimal}.

Consequently, we can enhance Eq.~\eqref{eq:final_P} to calculate the mismatch index $m$ as follows:
\begin{equation}
   m= \overline{p} -p_a =\sum_{r=1}^{\lfloor p_a (R+1) \rfloor} p^+(r,R) w (\frac{r}{R}-p_a) + \sum_{r=0}^{\lfloor p_a (R+1) \rfloor } p^-(r,R) w (\frac{r}{R+1}-p_a) - p_a.
   \label{eq:improve_m}
\end{equation}
Upon examining Fig.~\ref{fig:opitimal}, our findings indicate that in \emph{Hard} competition scenarios, individuals show a stronger tendency towards conservatism. This inclination is also present in the \emph{Easy} case, albeit to a lesser extent. In contrast, during \emph{Normal} competition scenarios, people tend to adopt a more adventurous approach. These observations provide valuable insights into understanding decision-making behavior across different levels of competitive intensity.

\section{Conclusion}
In conclusion, our study delves into the intricate dynamics of decision-making, specifically focusing on the job application process under the influence of limited information and intuitive decision-making. The efficiency of labor resource allocation is assessed through the mismatch index $m$, representing the disparity between the final application rate and the preset admission rate. Contrary to intuitive expectations, our simulation and analytical results reveal a counterintuitive trend: Under a naive strategy, as individuals acquire more information, the allocation becomes increasingly inefficient. This highlights the challenges associated with relying on intuition in the face of limited information, emphasizing the need for a more nuanced understanding of decision-making processes.

Additionally, our exploration uncovers a shift from despondent behavior to a bandwagon effect when the initial application rate exceeds the admission rate. This transition signifies the sensitivity of decision-making dynamics to the relationship between individual inclinations and the preset admission rate, especially in scenarios with global information. Furthermore, our study introduces a modification factor $w$ to characterize individuals' tendencies ranging from adventurous to conservative. This enhances the model's ability to capture a spectrum of individual behaviors in the real world. This approach bridges the gap between the model and real-world experiments, offering a more realistic depiction of decision-making processes.

In summary, our research offers valuable insights into resource allocation within the competitive job market, unraveling the intricacies arising from intuitive decision-making and the challenges posed by limited information. The counterintuitive findings emphasize the need for a nuanced approach to decision-making processes, particularly in competitive contexts. Looking ahead, several promising avenues for future research could further enhance our understanding and contribute to the broader field of decision science.

Firstly, factors such as individual personality traits, cultural influences, and situational variables may play pivotal roles in shaping decision-making strategies. Investigating these elements could provide a more comprehensive picture of the dynamics at play, offering insights into the varied approaches individuals adopt in competitive scenarios. Additionally, interventions aimed at enhancing decision-making efficiency in competitive environments warrant exploration. Developing training programs or decision support tools that mitigate the impact of cognitive biases and foster emotional intelligence could prove instrumental in improving decision outcomes. Last but not least, applying the model to different domains and contexts beyond job applications, such as college admissions, project funding, or investment decisions, could broaden its generalizability. Each domain presents its unique challenges and decision-making nuances, and adapting the model to these diverse scenarios would contribute to its versatility and practical relevance.

By embarking on these future research directions, we can refine existing models and develop novel strategies that optimize decision-making processes across a spectrum of competitive environments. The ultimate goal is to enhance the quality and efficiency of decisions made by individuals and organizations, fostering better outcomes in the competitive landscape.

~\\
\section*{Data Availability}
All original data can be found in the Supplementary Table.

\section*{Conflict of Interest Statement}
The authors declare no conflict of interest.

\section*{Acknowledgements}
We thank Zhehang Xu and Zhengming Xu who improved this paper with relevant comments. This work is supported by the National Natural Science Foundation of China (Grant No. 12205012, No. 61773148, and No. 52374013), and the Youth Talent Strategic Program (Grant No. 28705-310432106, No. 28704-310432104) from Beijing Normal University.

\bibliographystyle{unsrt}
\bibliography{ebb.bib}

\end{document}